\documentclass[12pt,a4paper]{article}

\usepackage{epsfig}
\usepackage{psfig}
\def\lsim{\raise0.3ex\hbox{$<$\kern-0.75em\raise-1.1ex\hbox{$\sim$}}}
\def\gsim{\raise0.3ex\hbox{$>$\kern-0.75em\raise-1.1ex\hbox{$\sim$}}}
\newcommand{\be}{\begin{equation}}
\newcommand{\ee}{\end{equation}}
\newcommand{\ba}{\begin{eqnarray}}
\newcommand{\ea}{\end{eqnarray}}

\def\spose#1{\hbox to 0pt{#1\hss}}
\def\ltapprox{\mathrel{\spose{\lower 3pt\hbox{$\mathchar"218$}}
 \raise 2.0pt\hbox{$\mathchar"13C$}}}
\def\gtapprox{\mathrel{\spose{\lower 3pt\hbox{$\mathchar"218$}}
 \raise 2.0pt\hbox{$\mathchar"13E$}}}
\def\NT{N_\tau}
\def\nt{\ifmmode\NT\else$\NT$\fi}
\def\NS{N_\sigma}
\def\ns{\ifmmode\NS\else$\NS$\fi}

\def\n{\noindent}

\begin{document}
\begin{titlepage} 
\thispagestyle{empty}

 \mbox{} \hfill BI-TP 2000/09\\
 \mbox{} \hfill hep-lat/0006023
\begin{center}
\vspace*{1.0cm}
{{\Large \bf Equation of state and Goldstone-mode effects  \\         
  of the three-dimensional $O(2)$ model\\}}\vspace*{0.5cm}
{\large J. Engels, S. Holtmann, T. Mendes and T. Schulze}\\ \vspace*{1.0cm}
\centerline {{\em Fakult\"at f\"ur Physik, Universit\"at Bielefeld, 
D-33615 Bielefeld, Germany}} \vspace*{0.5cm}
\protect\date \\ \vspace*{0.9cm}
{\bf   Abstract   \\ } \end{center} \indent
We investigate numerically the three-dimensional $O(2)$ model 
on $8^3$--$160^3$ lattices as a function of the magnetic field $H$.
In the low-temperature phase we verify the $H$-dependence of the 
magnetization $M$ induced by Goldstone modes and determine $M$ for
$V \rightarrow \infty$ on the coexistence line both by extrapolation
and by chiral perturbation theory. This enables us to calculate the 
corresponding critical amplitude. At $T_c$ the critical scaling behaviour  
of the magnetization as a function of $H$ is used to determine another
critical amplitude. In both cases we find negative corrections-to-scaling.
Our low-temperature results are well
described by the perturbative form of the model's magnetic equation 
of state, with coefficients determined nonperturbatively from our data.
The $O(2)$ scaling function for the magnetization is found to have  a 
smaller slope than the one for the $O(4)$ model. 
\vfill \begin{flushleft} 
\noindent{\rule[-.3cm]{5cm}{.02cm}} \\
\vspace*{0.2cm} 
E-mail: engels,holtmann,mendes,tschulze@physik.uni-bielefeld.de
\end{flushleft} 
\end{titlepage}

%%%%%%%%%%%%%%%%%%%%%%%%%%%%%%%%%%%%%%%%%%%%%%%%%%%%%%%%%%%%%%%%%%%%%%%%%%%%%%%%

\section{Introduction}

%%%%%%%%%%%%%%%%%%%%%%%%%%%%%%%%%%%%%%%%%%%%%%%%%%%%%%%%%%%%%%%%%%%%%%%%%%%%%%%%
$O(N)$ models are of general relevance to condensed matter 
physics and to quantum field theory, because many physical
systems exhibit a second-order phase transition with universal properties,
which may be derived analytically from $O(N)$ symmetric effective field 
theories \cite{Zinn}. A further aspect of these theories is remarkable. 
Due to the existence of massless Goldstone modes  
in $O(N)$ models with $N>1$ and dimension $2<d\leq4$ \cite{goldstone,madras} 
singularities are expected on the whole coexistence line $T<T_c, H=0$,   
in addition to the known critical behaviour at $T_c$. For $N=2$ and $d=3$
there exists a rigorous proof \cite{proof}. Numerically these
theoretical predictions have been recently confirmed by simulations of the
$3d$ $O(4)$ model \cite{EM}. In the same article a parametrization of the
equation of state including the Goldstone effect has been worked out and 
compared to perturbative predictions \cite{Brezin,WZ}. Here we want to extend
these numerical studies to the case of the $O(2)$ model. For the latter 
approximate representations of the equation of state have been derived, 
starting from high-temperature expansions \cite{Pel}. 

Our interest in the $O(2)$ model is moreover motivated by its relation 
to the staggered formulation of quantum chromodynamics (QCD) on the lattice.
At finite temperature QCD undergoes a chiral phase transition. For two 
degenerate light-quark flavors this transition is supposed to be of 
second order in the continuum limit and to belong to the same universality 
class as the $3d$ $O(4)$ model \cite{PW}-\cite{RW}. QCD lattice data have 
therefore been compared to the $O(4)$ scaling function, as determined 
numerically in \cite{Toussaint}. For staggered fermions the comparison to 
$O(4)$ is still not conclusive \cite{Edwin}-\cite{MILC}, but results for Wilson 
fermions \cite{wilson} seem to agree quite well with the predictions, though 
for the Wilson action the chiral symmetry is only restored in the continuum
limit. In the staggered formulation a part of the chiral symmetry is 
remaining even for finite lattice spacing, and that is $O(2)$. For the test 
of lattice QCD with staggered fermions it is therefore also important to know 
the corresponding universal $O(2)$ scaling function or equation of state.

The $O(2)$-invariant nonlinear $\sigma$-model (or $XY$ model), which we 
want to investigate in the following is defined by
\be
\beta\,{\cal H}\;=\;-J \,\sum_{<i,j>} {\bf S}_i\cdot {\bf S}_j
         \;-\; {\bf H}\cdot\,\sum_{i} {\bf S}_i \;,
\ee
where $i$ and $j$ are nearest-neighbour sites on a $d-$dimensional 
hypercubic lattice, and ${\bf S}_i$ is an 2-component unit vector 
at site $i$. It is convenient to decompose 
the spin vector ${\bf S}_i$ into a longitudinal (parallel to the magnetic 
field ${\bf H}$) and a transverse component 
\be
{\bf S}_i\; =\; S_i^{\parallel} {\bf \hat H} + {\bf S}_i^{\perp} ~.
\ee
The order parameter of the system, the magnetization $M$, is then the 
expectation value of the lattice average $S^{\parallel}$
of the longitudinal spin component
\be
M \;=\; <\!\frac{1}{V}\sum_{i} S_i^{\parallel}>\; =\; <  S^{\parallel}>~.
\ee
On finite lattices and $H=0$ system flips occur and lead to 
$<S^{\parallel}>=0$. Therefore one usually resorts to approximate order
parameter definitions, as e.g. $<|S^{\parallel}|>$ - see the discussion in 
Tapalov and Bl\"ote \cite{TB}. Here this is unnecessary since we always 
work at finite $H$.

There are two types of susceptibilities. The longitudinal 
susceptibility is defined as usual by the derivative of the magnetization, 
whereas the transverse susceptibility corresponds to the fluctuation 
of the lattice average ${\bf S}^{\perp}$ of the
transverse spin component
\ba
\chi_L\!\! &\!=\!&\!\! {\partial M \over \partial H}
 \;=\; V(<{ S^{\parallel}}^2>-M^2)~, \label{chil}\\
\chi_T\!\! &\!=\!&\!\! V < {{\bf S}^{\perp}}^2> 
~. \label{chit}
\ea
Both susceptibilities are predicted to diverge on the coexistence line 
$T<T_c, H=0$. The Goldstone singularities lead for all temperatures 
below $T_c$ for small $H$ to strong finite-size effects, which have 
been studied in the context of chiral perturbation theory for the 
three-dimensional $O(2)$ model already in \cite{Yone}. An explicit check on 
the $H$-dependence of the magnetization for $V\rightarrow\infty$
close to $H=0$ is however lacking. In comparison to the $O(4)$ model, 
we expect the Goldstone effects to be weaker, because we have only one 
transverse spin component in $O(2)$, not three like in $O(4)$. 
Another difference to $O(4)$ is that sizeable corrections to scaling 
appear in the $XY$ model \cite{Hase}. The determination of the 
universal equation of state of the three-dimensional $O(2)$ 
model, therefore requires a careful consideration of the corrections.
We could have avoided the problems with these corrections by working,
as it was done in Ref. \cite{Hase} at $\lambda=2.1$, instead of 
$\lambda=\infty$. For a better comparison to the $O(4)$ case \cite{EM}, 
we chose however to use the same model definition. Also there is then no 
need for an additional modulus update.

The plan of the paper is as follows. In the next section we 
briefly discuss the perturbative predictions for the 
magnetization and the susceptibilities at low temperatures, the form 
and the analytic results for the magnetic equation of state.
A more general review can be found in \cite{EM}.
Our numerical results are presented in Section \ref{section:results}, 
the equation of state is determined and parametrized
in Section \ref{section:sca_fun}. We close with a summary and our 
conclusions in Section \ref{section:conclusion}.

%%%%%%%%%%%%%%%%%%%%%%%%%%%%%%%%%%%%%%%%%%%%%%%%%%%%%%%%%%%%%%%%%%%%%%%%%%%%%%%%

\section{$\!\!$Perturbative Predictions and Equation of State}
\label{section:PT}

%%%%%%%%%%%%%%%%%%%%%%%%%%%%%%%%%%%%%%%%%%%%%%%%%%%%%%%%%%%%%%%%%%%%%%%%%%%%%%%%

For $T<T_c$ the system is in a broken phase, i.e.\ the
magnetization $M(T,H)$ attains a finite value $M(T,0)$ at $H=0$.
We explicitly assume here $H>0$. 
As a consequence the transverse susceptibility, which is 
directly related to the fluctuation of the Goldstone modes, 
diverges as $H^{-1}$ when $H\to0$ for all $T<T_c$. 
This is immediately clear from the identity \cite{Brezin}
\be
\chi_T \;=\; \frac{M(T,H)}{H}~,
\label{WI}
\ee
which is valid for all values of $T$ and $H$.
It is non-trivial that also the longitudinal susceptibility is diverging 
on the coexistence line for $2<d\leq4$. The leading term in the 
perturbative expansion for $2<d<4$ is $H^{d/2-2}$ \cite{goldstone,WZ}. 
The predicted divergence in $d=3$ is thus
\be
\chi_L(T<T_c,H)\;\sim\; H^{-1/2}~.
\label{chiL}
\ee
This is equivalent to an $H^{1/2}$-behaviour of the magnetization near 
the coexistence curve 
\be
M(T<T_c,H)\;=\;M(T,0)\,+\,c\,H^{1/2}~.
\label{magn}
\ee
An interesting question is whether this form of the magnetization
is compatible with the general 
Widom-Griffiths form of the equation of state \cite{Griffiths}
describing the critical behaviour in the vicinity of $T_c$. 
It is given by
\be
y\;=\;f(x)\;,
\label{eqstate}
\ee
where 
\be
y \equiv h/M^{\delta}, \quad x \equiv t/M^{1/\beta}.
\label{xy}
\ee
The variables $t$ and $h$ are the normalized 
reduced temperature $t=(T-T_c)/T_0$ and magnetic field $h=H/H_0$.
We take the usual normalization conditions 
\be
f(0) = 1, \quad f(-1) = 0~.
\label{normal}
\ee
The critical exponents $\delta$ and $\beta$ appearing in Eqs.~\ref{eqstate}
and \ref{xy} specify all the other critical exponents
\be
d\nu=\beta(1+\delta),\quad\gamma=\beta(\delta-1),\quad \nu_c=\nu/\beta\delta~.
\ee
Possible dependencies on irrelevant scaling fields
and exponents are however not taken into account in Eq.\ \ref{eqstate}, 
the function $f(x)$ is universal. Another way to express the 
dependence of the magnetization on $t$ and $h$ is
\be
M\;=\;h^{1/\delta} f_G(t/h^{1/\beta\delta})~,
\label{ftous}
\ee
where $f_G$ is a scaling function. This type of 
scaling equation is used for comparison to QCD lattice data.
The scaling forms in Eqs.\ (\ref{eqstate}) and (\ref{ftous}) are 
clearly equivalent.

The equation of state (\ref{eqstate}) has been derived by Br\'ezin
et al.\ \cite{Brezin} to order $\epsilon^2$ in the 
$\epsilon$-expansion, where $\epsilon=4-d$. 
The resulting approximation has been considered by Wallace and Zia \cite{WZ}
in the limit $x\to -1$, i.e.\ at $T<T_c$ and close to the coexistence 
curve. In this limit the function is inverted 
to give $x+1$ as a double expansion in powers of $y$ and $y^{d/2 - 1}$
\be
x+1\;=\; {\widetilde c_1} y \,+\, {\widetilde c_2} y^{d/2 - 1} \,+\,
         {\widetilde d_1} y^2 \,+\, {\widetilde d_2} y^{d/2} \,+\,
         {\widetilde d_3} y^{d-2} \,+\, \ldots \;.
\label{f_inv}
\ee
The coefficients ${\widetilde c_1}$, ${\widetilde c_2}$ and 
${\widetilde d_3}$ are then obtained from the general expression of
\cite{Brezin}. The above form is motivated by the $H$-dependence
in the $\epsilon$-expansion of $\chi_L$ at low temperatures \cite{WZ}.
In Section \ref{section:sca_fun} we propose a fit of our Monte Carlo data
to the perturbative form of the equation of state, using (\ref{f_inv}).

As for the large-$x$ limit (corresponding to $T>T_c$ and small $H$), 
the expected behaviour is given by Griffiths's analyticity condition
\cite{Griffiths}
\be
f(x) \;=\; \sum_{n=1}^{\infty} a_n\,x^{\gamma - 2(n-1)\beta} \;.
\label{Griffiths}
\ee
%%%%%%%%%%%%%%%%%%%%%%%%%%%%%%%%%%%%%%%%%%%%%%%%%%%%%%%%%%%%%%%%%%%%%%%%%%%%%%%%

\section{Numerical Results}
\label{section:results}

%%%%%%%%%%%%%%%%%%%%%%%%%%%%%%%%%%%%%%%%%%%%%%%%%%%%%%%%%%%%%%%%%%%%%%%%%%%%%%%%
Our simulations are done on three-dimensional lattices with periodic 
boundary conditions and linear extensions $L=8-160$ using the cluster 
algorithm of Ref.\ \cite{clust}. The value $J_c = 0.454165(4)$, 
obtained in simulations of the zero-field model \cite{Spain}, is used in 
the following. We have simulated at increasingly larger values of $L$ at 
fixed values of $J=1/T$ (i.e.\ at fixed temperature $T$) and $H$ in order 
to eliminate finite-size effects.
 
\begin{figure}[htb]
\begin{center}
   \epsfig{bbllx=127,bblly=264,bburx=451,bbury=587,
        file=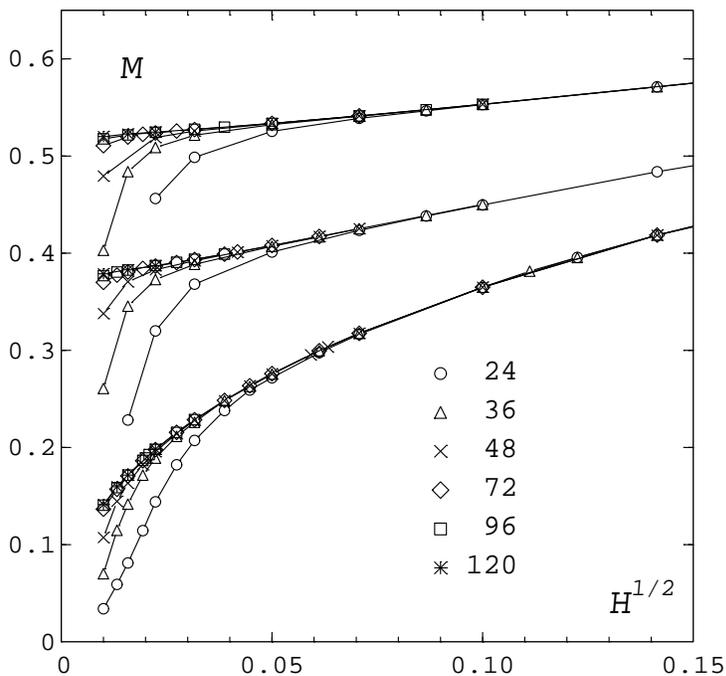,width=84mm}
\end{center}
\caption{The magnetization as a function of $H^{1/2}$ in the 
low-temperature region for fixed $J=0.5$, 0.47 and $J_c$, starting 
with the highest curve for various $L$.} 
\label{fig:mroot}
\end{figure}

As an example we show
in Fig.\ \ref{fig:mroot} part of our data for the magnetization 
for low temperatures, at $J=0.5,~0.47$, and at $J_c$, plotted versus 
$H^{1/2}$. The picture is rather similar to the one obtained in $O(4)$ 
\cite{EM}: strong finite-size effects appear for small $H$ and persist 
as one moves away from $T_c$, the results from the largest lattices are 
at first sight linear in $H^{1/2}$, as predicted by Eq.\ \ref{magn}, 
below but not at $T_c$. A closer look at the curves reveals however 
subtle differences compared to the $O(4)$ case, which are confirmed by 
the attempt to determine the value of $M(T,H=0)$ by a linear fit in 
$H^{1/2}$. Very close to $H=0$ the fixed temperature curves become 
slighty flatter, leading to a higher value for $M(T,0)$ than expected 
from the data at larger $H$ values. This behaviour is more pronounced 
close to $T_c$ than at lower temperatures. The determination of the 
magnetization in the thermodynamic limit ($V\rightarrow \infty$) on the 
coexistence line requires therefore more precise data and smaller values 
of $H$ than for the $O(4)$ model. In order to extrapolate the data to 
$H\rightarrow 0$ and $V\rightarrow \infty$ we apply two different 
strategies. The first is to extend the linear form in $H^{1/2}$, 
Eq.\ \ref{magn}, to a quadratic one \cite{goldstone}
\be
M(T<T_c,H)\;=\;M(T,0)\,+\,c_1\,H^{1/2}+\,c_2\,H~,
\label{magn2}
\ee
and to fit the data from the largest lattices, which we assume to represent
data on an infinite volume lattice, to this form. The second way to find
$M(T,0)$ is just opposite to the first. Here we exploit the $L$ or volume
dependence at fixed $J$ and fixed small $H$ to determine via chiral 
perturbation theory (CPT) the magnetization $\Sigma$ of the continuum 
theory for $V\rightarrow \infty,~H=0$, which is related to 
%----------------------------------------------------------------------------
%----------------------------------------------------------------------------
$M(T,0)$ by 
\be
M(T,0)\;=\;{\Sigma \over \sqrt{J}}~.
\label{magn3}
\ee
\n According to the $\epsilon$-expansion of CPT 
the quantity $\Sigma$ and the Goldstone boson decay constant $F$  
are the only parameters which determine the finite-size corrections 
to the partition function to order $1/L^2$ \cite{HL}. We summarize the 
relevant formulae for the three-dimensional $XY$ model from Ref.\ \cite{Yone}.
In our notation the magnetization is
\be
M\;=\;{\Sigma \over \sqrt{J}}u[ \rho_1 \eta+2\rho_2 \alpha^2]~,
\label{magn4}
\ee
where $u=\rho_1 \Sigma H V/ \sqrt{J}$, $\alpha=1/(F^2L)$ and
\be
\rho_1 = 1+{1\over 2}\beta_1 \alpha + {1 \over 8}(\beta_1^2-2\beta_2)
\alpha^2~;\quad \rho_2={1\over 4} \beta_2~.
\ee
For a symmetric three-dimensional box $\beta_1=0.225785$ and 
$\beta_2=0.010608$. The parameter $\eta$ is given in terms of 
modified Bessel functions as
\be 
\eta\;=\; {I_1(u) \over uI_2(u)}~.
\ee
By construction the $\epsilon$-expansion is only applicable in a range 
where $m_{\pi}L\lsim 1$ which translates into the condition 
\be
H{\Sigma \over \sqrt{J}}\; \lsim \;\left({F \over L}\right)^2~.
\label{cond}
\ee
\n In Fig.\ \ref{fig:mrhall} we show the data for the magnetization as a 
function of $H^{1/2}$ for six fixed values of $J$ in the low-temperature 
phase. We observe a remarkable coincidence of the fits according to Eq.\ 
\ref{magn2} with the CPT results at $H=0$. Details of
the fits are presented in Table \ref{tab:mfits}. The $\chi^2$ per degree of 
freedom for the first fit type is in the range 0.01-0.5~. 
The corresponding numbers for the CPT fits are of the order of 1.  
Like for the $O(4)$ model we see in Fig.\ \ref{fig:mrhall} that the region
where $M$ is linear in $H^{1/2}$ shrinks when $T$ approaches $T_c$,
the value of $c_2$ in Table \ref{tab:mfits} increases correspondingly.
At $J=0.460$ the fit parabola coincides only on a small piece with the largest
volume data (here we simulated even on $L=160$ lattices), the errors on $c_1$
and $c_2$ become large, though the value at $H=0$ agrees again nicely with the 
CPT result.
We have further convinced ourselves that $M(T,0)$ can also be obtained
from calculations at $H=0$ with the modulus definition. The approach to the
thermodynamic limit is here from above.

%------------------------------------------------------------------------
\setlength{\unitlength}{1cm}
\begin{picture}(13,7)
\put(-0.1,0){
   \epsfig{bbllx=127,bblly=264,bburx=451,bbury=587,
       file=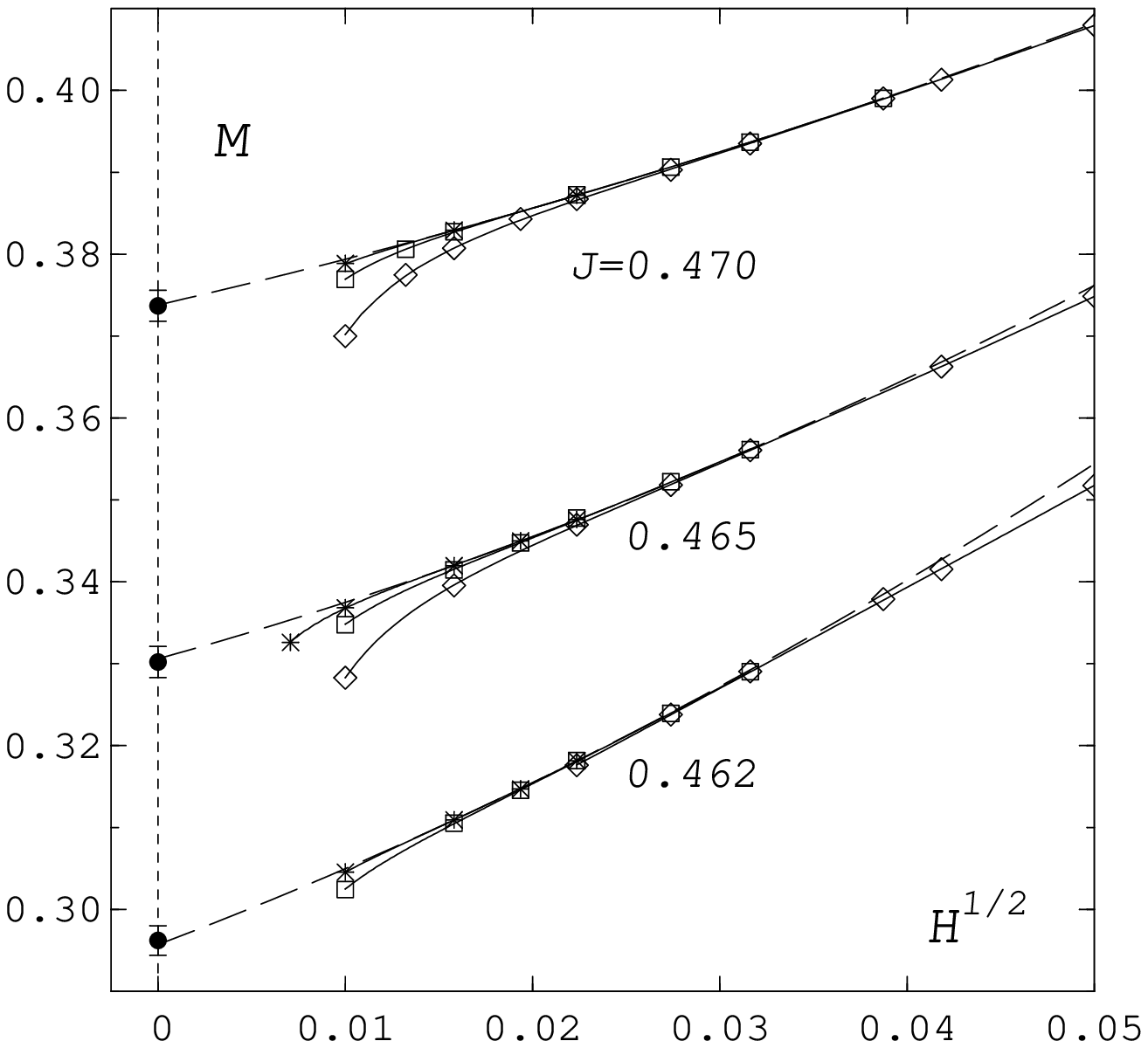, width=65mm}
          }
\put(7.5,0){
   \epsfig{bbllx=127,bblly=264,bburx=451,bbury=587,
       file=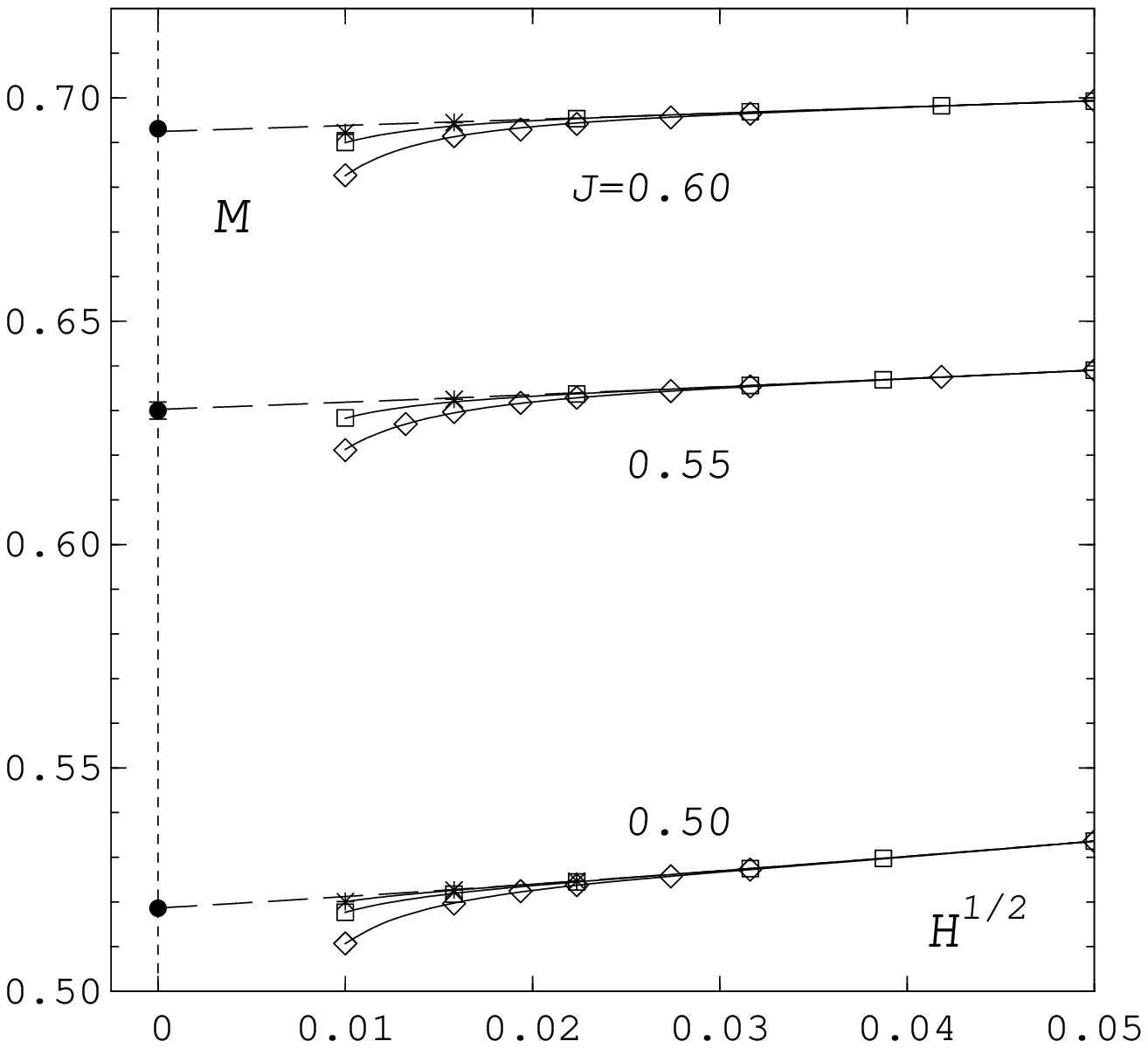, width=65mm}
          }
\end{picture}
\begin{figure}[h!]
\caption{The magnetization vs.\ $H^{1/2}$ in the low-temperature region
for fixed $J=0.47$, 0.465, 0.462 (left plot) and $J=0.6$, 0.55, 0.5 (right
plot) and different $L$ (notation as in Fig.\ref{fig:mroot}). The filled
circles are the results from CPT, the dashed lines the fits according to 
Eq.\ \ref{magn2}. The solid lines connecting the data are from reweighting.}
\label{fig:mrhall}
\end{figure}
%----------------------------------------------------------------------------

%----------------------------------------------------------------------------
\begin{table}[ht]
\begin{center}
  \begin{tabular}{|c|cccc|cr|}
    \hline
$J=1/T$ & $M(T,0)$ & $c_1$ & $c_2$ & $10^4\cdot H$ & $\Sigma/\sqrt{J}$ & $L~~$
\\ \hline
 0.460 &0.2724(27) & 0.53(30)  & 21.9(80) & 2-5 & 0.2720(27)& 12-40\\  
 0.462 &0.2957(03) & 0.865(30) & 6.18(82) & 2-5 & 0.2962(18)& 12-36\\ 
 0.465 &0.3306(02) & 0.633(17) & 5.58(36) & 3-10& 0.3302(19)& 12-40\\
 0.470 &0.3738(01) & 0.526(03) & 3.24(05) & 3-15& 0.3737(19)&  8-40\\
 0.500 &0.5186(01) & 0.244(02) & 1.11(02) & 8-75& 0.5186(12)& 16-48\\
 0.550 &0.6303(01) & 0.157(01) & 0.38(01) & 10-50&0.6300(19)& 12-56\\
 0.600 &0.6925(01) & 0.133(01) & 0.09(01) & 16-75&0.6931(07)&  8-64\\ \hline
  \end{tabular}
\end{center}
\caption{Parameters of the fit of $M(T,H)$ to Eq.\ \ref{magn2} in the 
$H$-range of column 5 and the results of the CPT fit for $H=0.0001$ in
the $L$-range of column 7.}

\label{tab:mfits}
\end{table}
%----------------------------------------------------------------------------
\n In the neighbourhood of the critical temperature the results for 
$M(T,0)$ should show the usual critical behaviour. Contrary to the $O(4)$
case we expect here a sizeable correction to the leading scaling behaviour
\cite{Hase}. In order to determine the critical amplitude $B$ we 
therefore make the following ansatz 
\be
M(T\ltapprox T_c,H=0) \;=\;B(T_c - T)^{\beta}[1\;+\;b_1(T_c-T)^{\omega \nu}
\;+\;b_2(T_c-T)]~.
\label{cocrit}
\ee
\n As an input we take the critical exponents from Ref.\ \cite{Hase} 
\be
\beta\;=\;0.3490(6)~,\quad \nu\;=\;0.6723(11)~,\quad \omega\;=\;0.79(2)~.
\label{Hasex}
\ee
A fit of all points of Table \ref{tab:mfits}, apart from the one at
$J=0.46$, using the form (\ref{cocrit}) leads to the result 
$B=0.945(5),~b_1=-0.053(23)$ and $b_2=-0.098(23)$. In Fig.\ \ref{fig:mcoex}a
we show this fit and also the leading term separately.

%------------------------------------------------------------------------

\setlength{\unitlength}{1cm}
\begin{picture}(13,7)
\put(-0.1,0){
   \epsfig{bbllx=127,bblly=264,bburx=451,bbury=587,
       file=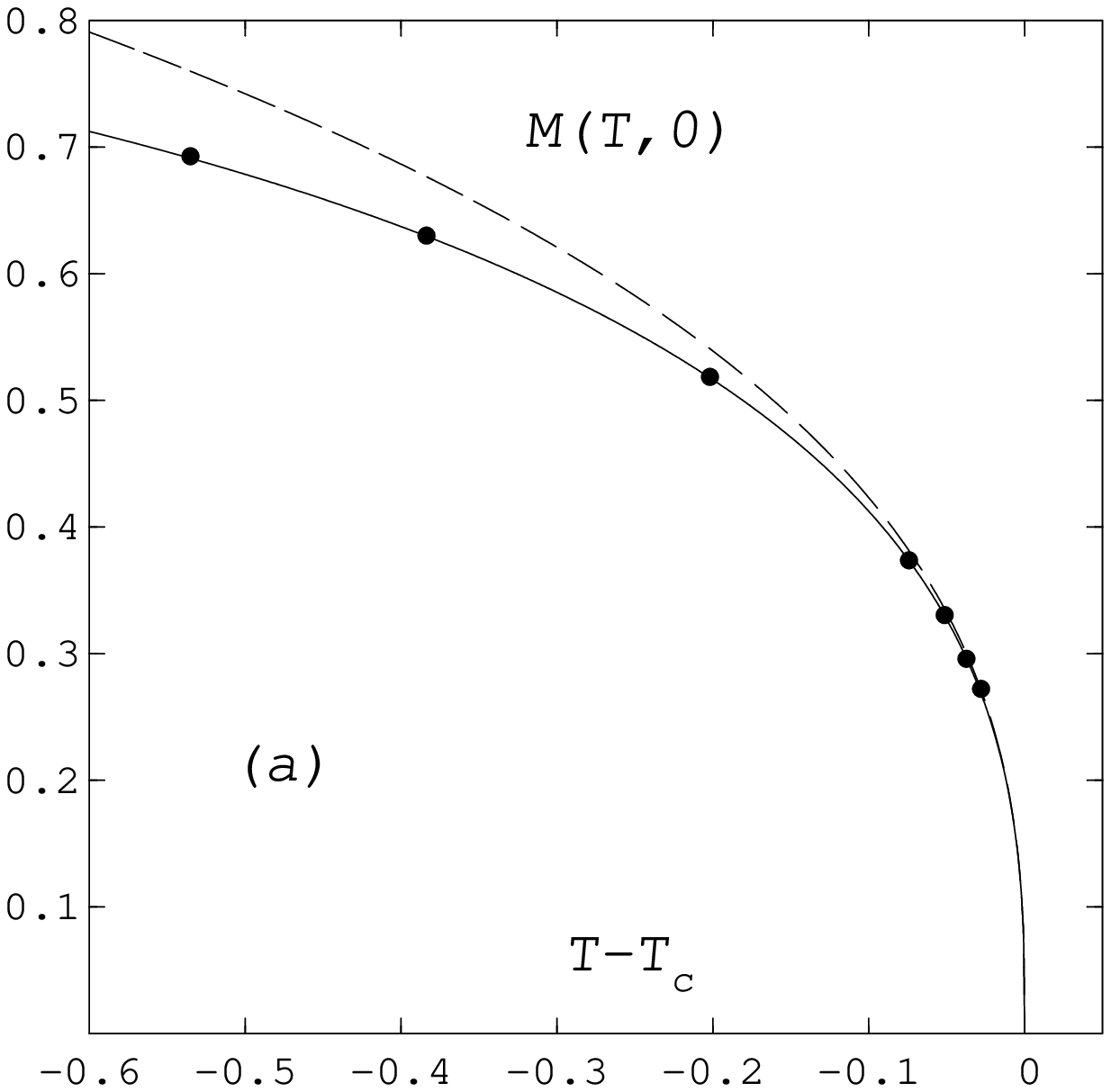, width=65mm}
          }
\put(7.5,0){
   \epsfig{bbllx=127,bblly=264,bburx=451,bbury=587,
       file=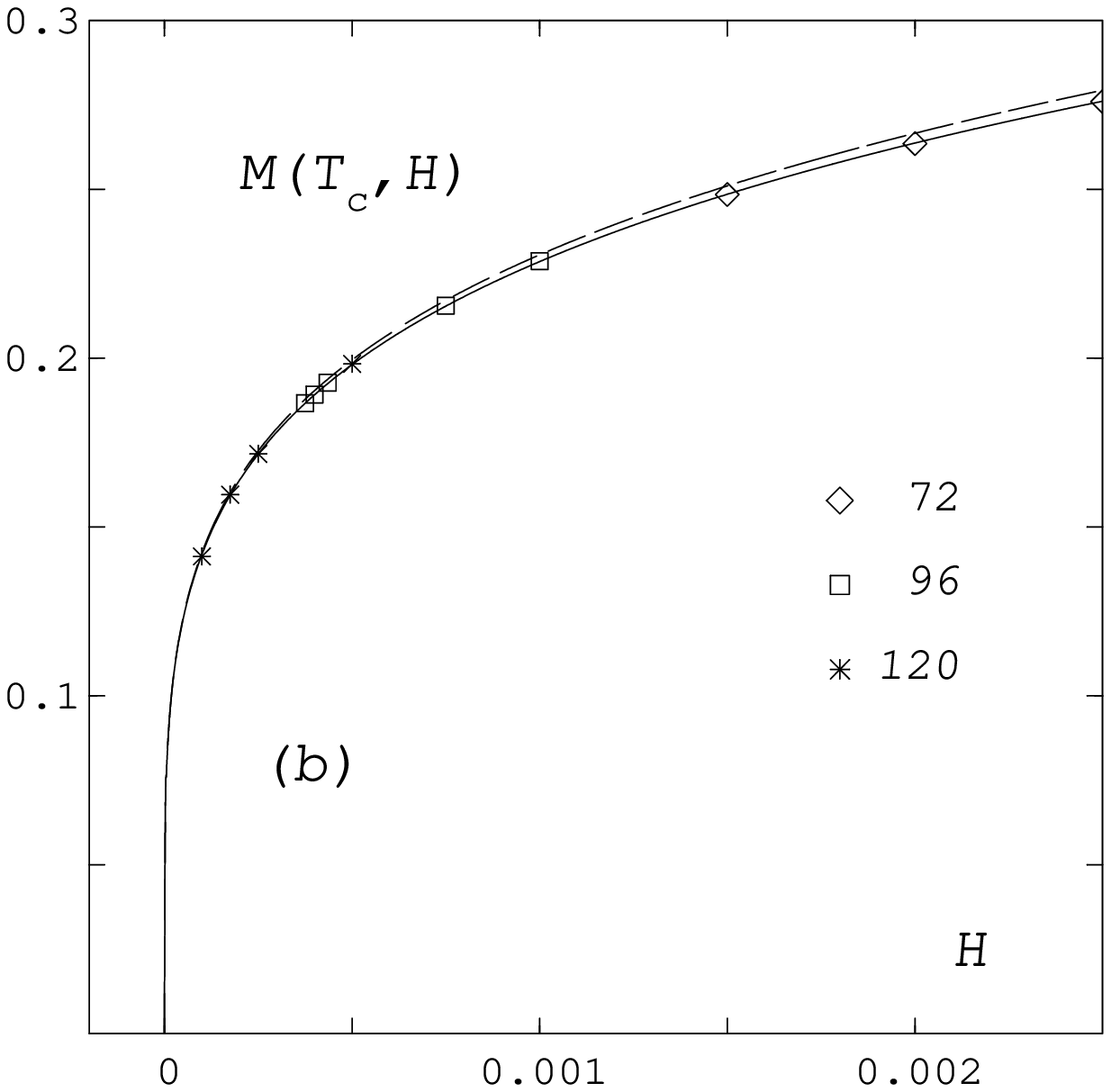, width=65mm}
          }
\end{picture}
\begin{figure}[h!]
\caption{(a) The magnetization on the coexistence line versus $T-T_c$, 
the points are from Table \ref{tab:mfits}, the solid line is the fit\
(\ref{cocrit}), the dashed line its leading part.\break (b) The magnetization 
at $T_c$ as a function of $H$, the line is the fit\ (\ref{hcrit}),
the dashed line the leading part.}
\label{fig:mcoex}
\end{figure}
%----------------------------------------------------------------------------

\n As the critical point is reached the $H$-dependence of the magnetization 
changes to satisfy critical scaling. We therefore fit the data from the 
largest lattice sizes at $T_c$ to the form
\be
M(T_c,H) \;=\;d_c H^{1/\delta}[ 1 + d_c^1 H^{\omega \nu_c}]~.
\label{hcrit}
\ee
A further term proportional to $H$ is unnecessary, because the corrections
to scaling are much smaller here, than on the coexistence line. The largest 
$L$ data can be fitted very well with the ansatz (\ref{hcrit}) and the 
critical exponents as input, as can be seen in Fig.\ \ref{fig:mcoex}b. We
find $d_c=0.978(2)$ and $d_c^1=-0.075(5)$.

\begin{figure}[htb]
\begin{center}
   \epsfig{bbllx=127,bblly=264,bburx=451,bbury=587,
        file=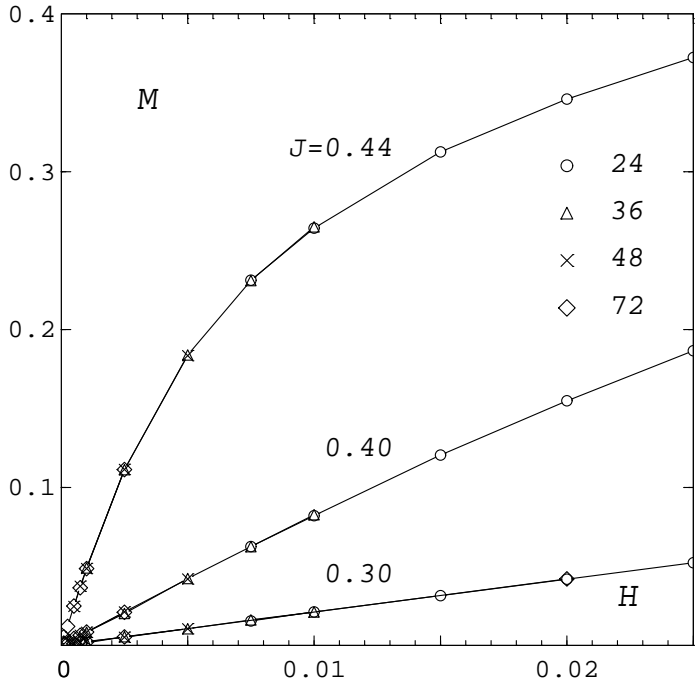,width=84mm}
\end{center}
\caption{The magnetization as a function of $H$ in the 
high-temperature region for fixed $J=0.44$, 0.4 and 0.3, starting 
with the highest curve for various $L$.} 
\label{fig:mhigh}
\end{figure}

\n We have performed in addition some simulations in the high temperature 
phase of the model. Here the finite size effects are rather small. In Fig.\
\ref{fig:mhigh} we show the results. One observes that with increasing $T$
(decreasing $J$) the region of linear dependence on $H$ of the magnetization 
becomes larger.    
%----------------------------------------------------------------------------
%----------------------------------------------------------------------------

%%%%%%%%%%%%%%%%%%%%%%%%%%%%%%%%%%%%%%%%%%%%%%%%%%%%%%%%%%%%%%%%%%%%%%%%%%%%%%%%
\section{Determination of the Equation of State}
\label{section:sca_fun}

%%%%%%%%%%%%%%%%%%%%%%%%%%%%%%%%%%%%%%%%%%%%%%%%%%%%%%%%%%%%%%%%%%%%%%%%%%%%%%%%

In the last section we derived from our Monte Carlo data the magnetization
in the thermodynamic limit. At temperatures below $T_c$ this was achieved 
for the larger $H$ values by simply performing simulations on lattices with 
increasing size until all volume dependence was gone. For the small $H$ 
region down to $H=0$ we took then advantage of the Goldstone effect to 
extrapolate our data and confirmed this by the CPT results for $\Sigma$. In the 
high-temperature phase we reached the thermodynamic limit already on lattices
with $L\le72$ for all $H$ values.    

\begin{figure}[htb]
\begin{center}
   \epsfig{bbllx=127,bblly=264,bburx=451,bbury=587,
        file=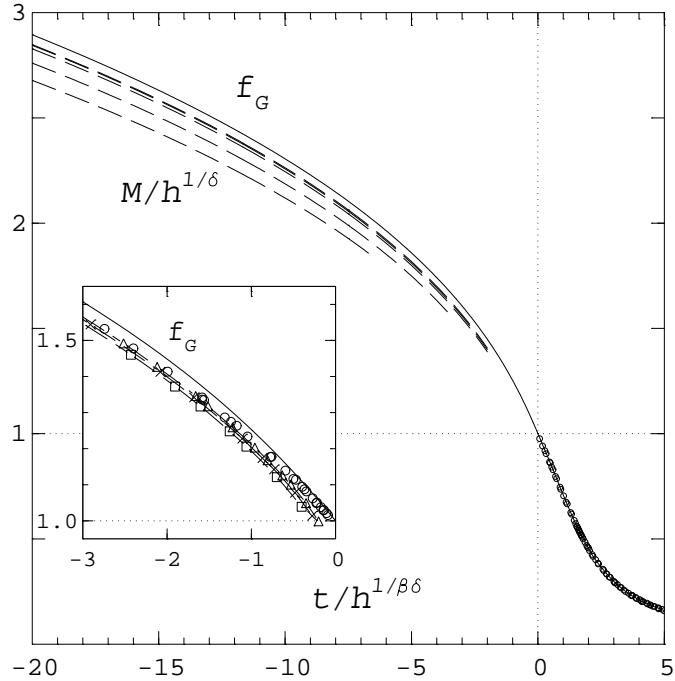,width=84mm}
\end{center}
\caption{The scaling function $f_G$, Eq.\ (\ref{ftous}),
(solid line). Also shown are the results for $M/h^{1/\delta}$ at
fixed values of $J=0.55,0.50,
0.47,0.465$ and 0.462 (dashed lines) which were used for the extrapolation,
starting with the lowest curve. The circles are single data points. 
The inset shows the low-temperature region close to $T_c$, with data 
for $J\le 0.46$ (circles), 0.462 (triangles), 0.465 (crosses) and 0.47 
(squares).} 
\label{fig:fgscale}
\end{figure}

\n The two critical amplitudes $B$ and $d_c$ can now be 
utilized to normalize temperature and magnetic field according to Eq.\
(\ref{normal}). We find from our fits (\ref{cocrit}) and (\ref{hcrit})
\be
T_0 = B^{-1/\beta} = 1.18(2)~,~{\rm and}~H_0 = d_c^{-\delta} = 1.11(1)~. 
\ee 
As mentioned already, the equation of state does not account for possible 
corrections to scaling. A more general form of Eq.\ (\ref{ftous}) is 
\be 
Mh^{-1/\delta}\; =\; \Psi (th^{-1/\beta\delta}, h^{\omega\nu_c})~.
\ee
Expanding the function $\Psi$ in $h^{\omega\nu_c}$ leads to
\be
Mh^{-1/\delta}\; =\; f_G(th^{-1/\beta\delta}) + h^{\omega\nu_c}
 f_G^{(1)}(th^{-1/\beta\delta}) + \cdots~.
\ee
In order to obtain the scaling function $f_G$ we therefore perform
quadratic fits to our data in $h^{\omega\nu_c}$ at fixed values of 
$th^{-1/\beta\delta}$ in the low-temperature region, where the 
corrections are strong. Fortunately, the corrections are very 
small in the high-temperature region and the data scale directly.
Altogether data with $H \leq 0.0075$ and $0.43 \le J \le 0.55$
were used for this purpose. In Fig.\ \ref{fig:fgscale} we have plotted 
the uncorrected and the final results for the scaling function $f_G$.
%----------------------------------------------------------------------------
%----------------------------------------------------------------------------

\n Like in the case of $O(4)$ \cite{EM} we want to describe the low-temperature
part of the equation of state by its perturbative form as discussed in 
Section \ref{section:PT}, but with nonperturbative coefficients. We then want, 
again as in Ref.\ \cite{EM}, to interpolate this result with a fit to the 
large-$x$ form (\ref{Griffiths}). The variables $x$ and $y$ are simply related 
to the scaling function $f_G$ and its argument by
\be
 y\;=\;f_G^{-\delta} \;, \quad  
x \;=\; (t/h^{1/\beta\delta})\, f_G^{-1/\beta}\;.
\ee

\n We first perform a fit to $x$ in the interval $[-1,\,1.4]$ (which 
corresponds to $y\in[0,3]$) with the three leading terms in (\ref{f_inv}) 
\be
x_1(y)+1 \;=\; ({\widetilde c_1} \,+\, {\widetilde d_3})\,y \,+\,
             {\widetilde c_2}\,y^{1/2} \,+\, 
             {\widetilde d_2}\,y^{3/2} \;.
\label{PTform}
\ee
In the fit we require ${\widetilde d_2}=1-({\widetilde c_1} \,+\,
 {\widetilde d_3}\,+\,{\widetilde c_2})$ to fix the normalization $y(0)=1$.
We obtain
\be
{\widetilde c_1} + {\widetilde d_3} \,=\, 0.352(30) \;,\quad
{\widetilde c_2} \,=\, 0.592(10) \;.
\label{result}
\ee
\n The fit describes the corrected scaling function at $T<T_c$ and 
and the direct data in the high temperature phase 
up to $x\approx 1.7$. This confirms, as in $O(4)$, that the expression 
(\ref{f_inv}) is valid also away from $x\approx -1$.
Our coefficients can be compared to those calculated perturbatively for
$N=2$ in Ref.\ \cite{WZ}
\be
{\widetilde c_1} + {\widetilde d_3} \,=\, 0.818 \;,\quad
{\widetilde c_2} \,=\, 0.229 \;.
\label{expan}
\ee
The result (\ref{result}) is much closer to the coefficients found in $O(4)$
\cite{EM} than to the $\epsilon$-expansion results (\ref{expan}), though the
Goldstone effect is decreasing somewhat with decreasing $N$: ${\widetilde c_2}$
and the $x$ region where the ansatz (\ref{PTform}) is valid
are a little smaller.
   
%------------------------------------------------------------------------
\setlength{\unitlength}{1cm}
\begin{picture}(13,7)
\put(-0.3,0){
   \epsfig{bbllx=127,bblly=264,bburx=451,bbury=587,
       file=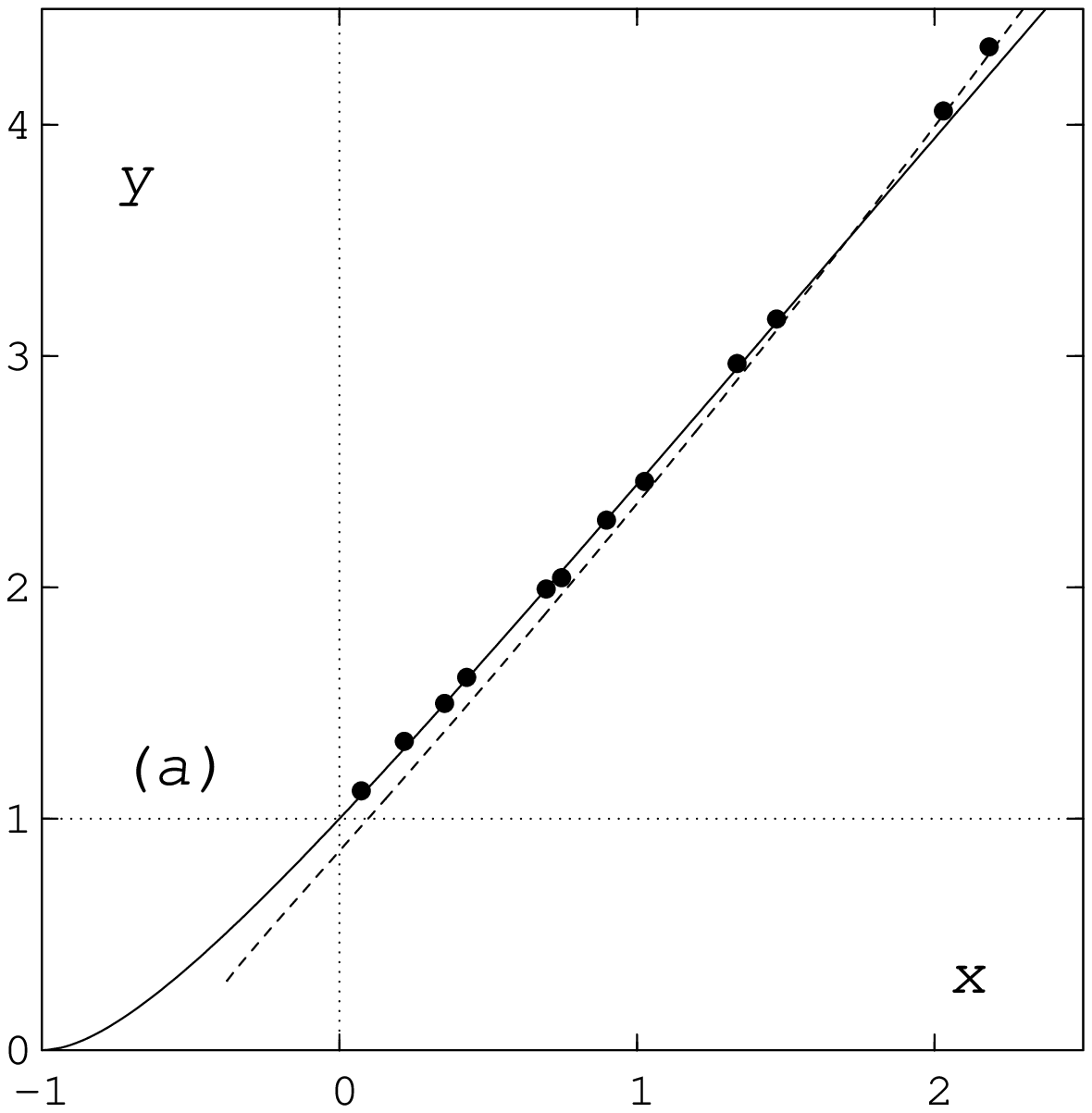, width=67mm}
          }
\put(7.3,0){
   \epsfig{bbllx=127,bblly=264,bburx=451,bbury=587,
       file=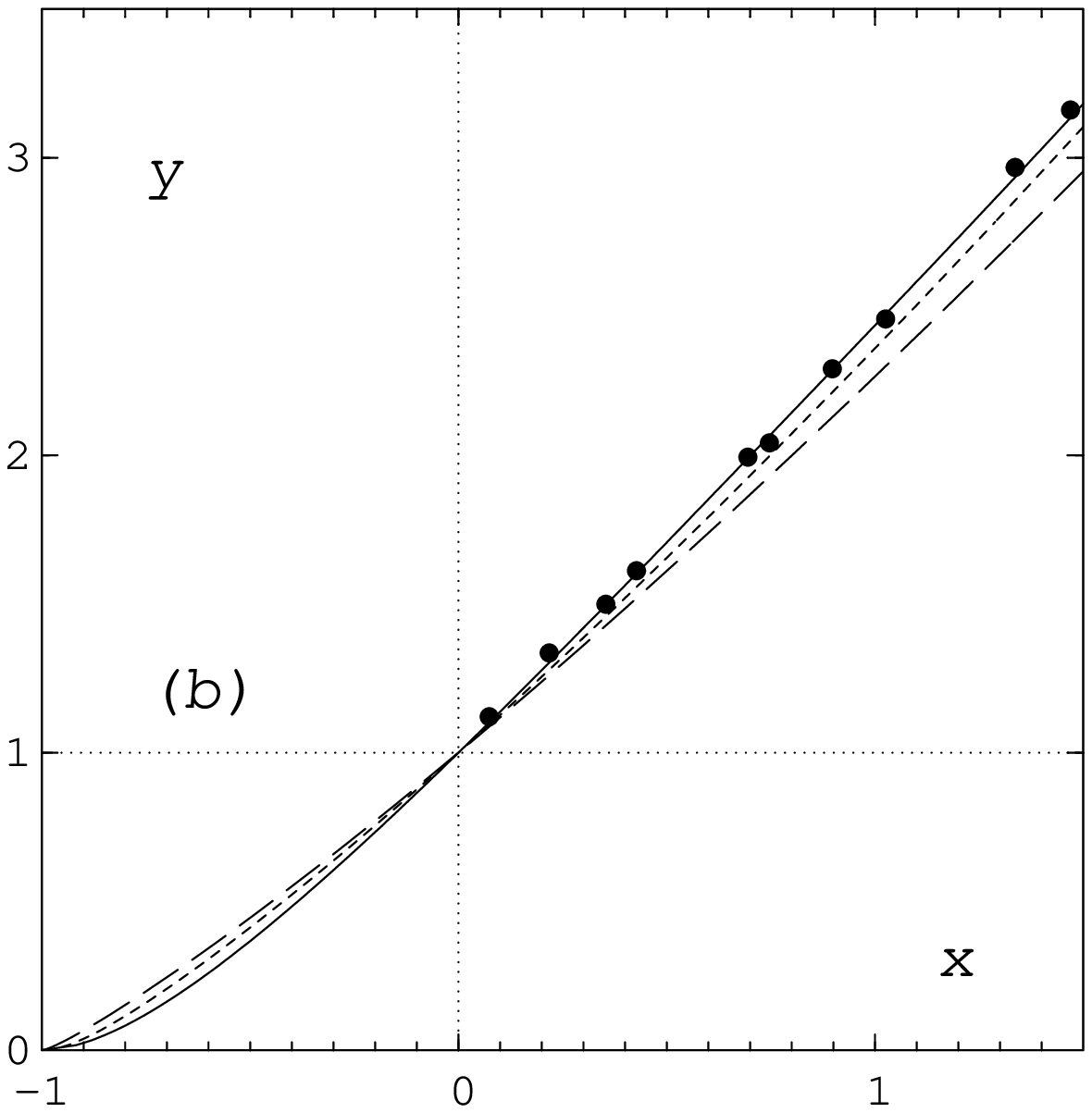, width=67mm}
          }
\end{picture}
\begin{figure}[h!]
\caption{(a) The function $y=f(x)$ from fits at small $x$ (solid line) 
and at large $x$ (dashed line). (b) The interpolation (\ref{totalfit}) 
for $f(x)$ (solid line), and the parametrizations (A) (short dashs) and
(B) (long dashs), $n=1$ of $f(x)$ from Ref.\ \cite{Pel}. In both parts 
we show high-temperature data ( filled circles).}
\label{fig:yfx}
\end{figure}
%----------------------------------------------------------------------------

\begin{figure}[ht]
\begin{center}
   \epsfig{bbllx=127,bblly=264,bburx=451,bbury=587,
        file=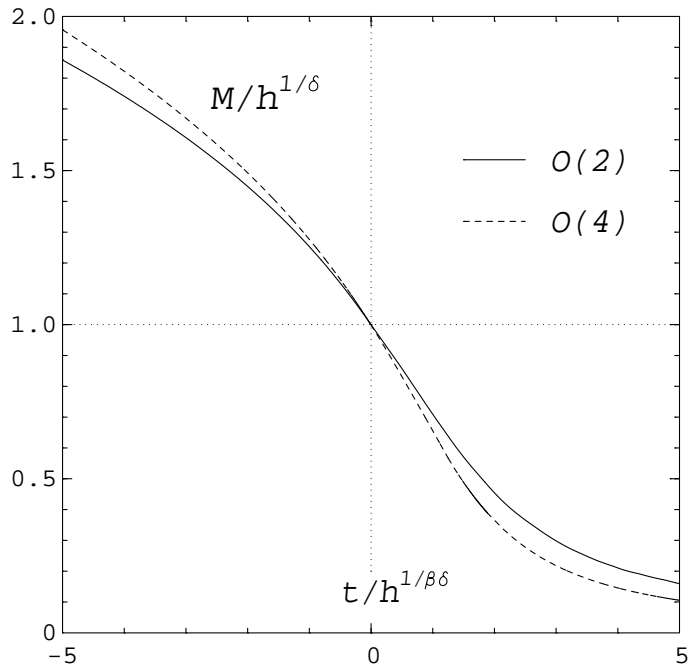,width=82mm}
\end{center}
\caption{The scaling function $f_G=M/h^{1/\delta}$ for the $O(2)$ model
(solid line) and the $O(4)$ model (dashed line).}
\label{fig:scafun}
\end{figure}

\n For large $x$ we have done a 2-parameter fit
of the behaviour (\ref{Griffiths}), in the corresponding form for
$x$ in terms of $y$
\be
x_2(y)\;=\; a\, y^{1/\gamma} \,+\, b\,y^{(1-2\beta)/\gamma}~.
\label{highx}
\ee
Considering data points with $y$ in the interval [4,23000] we find
\be
a \,=\, 1.2595(30)\;, \quad  b \,=\, -1.163(20) \;.
\ee
Expression (\ref{highx}) describes the data for $x > 1.5$.
The small- and large-$x$ curves cover the whole range
of values of $x$ remarkably well. The two curves overlap around 
$y \approx 3.5$. This is shown in Fig.\ \ref{fig:yfx}a.
We therefore interpolate them smoothly with the ansatz
\be
x(y) \;=\; x_1(y)\,\frac{y_0^6}{y_0^6 + y^6} \,+\,
           x_2(y)\,\frac{y^6}{y_0^6 + y^6}~,
\label{totalfit}
\ee
where $y_0=3.5$. In Fig.\ \ref{fig:yfx}b we compare our interpolation
(\ref{totalfit}) of the equation of state to the high-temperature data
and to two parametric forms for $f(x)$ obtained from the high-temperature 
expansion in Ref.\ \cite{Pel}.
According to the authors \cite{aua}, the difference between these two
curves gives an idea of the uncertainty of their approach. For large $x$ this
difference can be traced back to the uncertainty in the universal amplitude 
ratio $R_{\chi}$ (Eq. (49) of Ref.\ \cite{Pel}), which was determined in 
\cite{Pel} to $R_{\chi}=1.4(1)$, that is with a 7\% error. In fact, $R_{\chi}$
is related to the quantity $a$ of our large-$x$ parametrization by
\be
R_{\chi}\;=\; a^{\gamma}\;=\;1.356(4)~,
\ee
which is in agreement with the result from Ref.\ \cite{Pel}.
Their low temperature parameter $c_f$ may be calculated from our value
of ${\widetilde c_2}$ and amounts to $c_f=2.85(7)$.
Finally we show in Fig.\ \ref{fig:scafun} the scaling function $f_G$ 
obtained parametrically from $x(y)$ in (\ref{totalfit}) and the 
corresponding scaling function for $O(4)$. As can be seen from 
Fig.\ \ref{fig:scafun}, the dependence on $t/h^{1/\delta\beta}$ is
similar, but the $O(2)$ curve is flatter than the one for $O(4)$.
\smallskip
%----------------------------------------------------------------------------
%----------------------------------------------------------------------------

%%%%%%%%%%%%%%%%%%%%%%%%%%%%%%%%%%%%%%%%%%%%%%%%%%%%%%%%%%%%%%%%%%%%%%%%%%%%%%%%
\section{Summary and Conclusions}
\label{section:conclusion}
%%%%%%%%%%%%%%%%%%%%%%%%%%%%%%%%%%%%%%%%%%%%%%%%%%%%%%%%%%%%%%%%%%%%%%%%%%%%%%%%

We have simulated the three-dimensional $O(2)$ model on cubic lattices
as a function of the magnetic 
field $H$ and the temperature $T$. From the behaviour of the measured
magnetization $M$ below the critical temperature $T_c$ and close to the
coexistence line ($H=0$), we could clearly verify the Goldstone-mode effects.
This was done in two independent ways which led to the same values $M(T<T_c,
H\rightarrow 0)$ for the magnetization in the thermodynamic limit $V
\rightarrow \infty$ : on one hand we were able to profit by the observed 
Goldstone behaviour to extrapolate our data to $H\to0$ and on the other
hand this result was confirmed from the finite-size dependence induced 
by the Goldstone modes, using chiral perturbation theory.

The $M$ values on the coexistence line were subsequently used to calculate
the critical amplitude $B$ of the magnetization. Not unexpectedly, we found 
here strong corrections to scaling. On the critical line $T=T_c$ we computed
then the second critical amplitude $d_c$ of the magnetization. Here, the 
corrections to scaling are less pronounced ; in the high-temperature phase
they are unimportant. 

The problem with corrrections to scaling appeared again in the determination
of the equation of state in the low-temperature phase. By generalizing the
scaling equation to include possible corrections, we were able to derive
the universal form of the equation of state for the $O(2)$ model from our 
data. This form was then analyzed in a similar manner as it was done for
the $O(4)$ model in Ref.\ \cite{EM}. In particular, we found again an efficient
parametrization of the equation of state in the low $x$ region with the 
perturbative form (\ref{PTform}), which is based on expansion (\ref{f_inv}) 
by Wallace and Zia \cite{WZ}. The effect due to the Goldstone modes is nearly
as large as in the $O(4)$ case, as can be seen from the coefficient 
${\widetilde c_2}$ and much larger than the prediction from the   
$\epsilon$-expansion. 

Like for $O(4)$ we found a very good large-$x$ fit to the 
high-temperature data. The coefficient $a$ obtained from the fit implies
a value of $R_{\chi}=1.356(4)$, in agreement with parametrizations of the 
equation of state by Campostrini et al.\ \cite{Pel}. A further, indirect check
of the equation of state is the computation of the universal ratio $A_+/A_-$. 
This requires however an integration of the magnetization with respect to $h$ 
and derivatives with respect to $t$. We shall consider this in the 
near future.

Upon interpolation with the low-$x$ curve, 
a complete description for the equation of state is obtained, which
can be plotted parametrically also for the scaling function.
In Ref.\ \cite{MILC} the lattice QCD data of the MILC Collaboration
for $N_{\tau}=4$ were compared to the $O(4)$ scaling function. The test failed
because the data were indicating a steeper scaling function. Since
the $O(2)$ scaling function is flatter than the one for $O(4)$, the situation
will be worse there. A way out may be the comparison to finite-size-scaling
functions, since lattice QCD is presumably far from the thermodynamic limit.
We are investigating this currently.

%%%%%%%%%%%%%%%%%%%%%%%%%%%%%%%%%%%%%%%%%%%%%%%%%%%%%%%%%%%%%%%%%%%%%%%%%%%%%%%%
\vskip 0.2truecm
\noindent{\Large{\bf Acknowledgements}}

%%%%%%%%%%%%%%%%%%%%%%%%%%%%%%%%%%%%%%%%%%%%%%%%%%%%%%%%%%%%%%%%%%%%%%%%%%%%%%%%

\n We thank Attilio Cucchieri for helpful suggestions and comments.
This work was supported by the Deutsche Forschungs\-ge\-meinschaft
under Grant No.\ Ka 1198/4-1.
%%%%%%%%%%%%%%%%%%%%%%%%%%%%%%%%%%%%%%%%%%%%%%%%%%%%%%%%%%%%%%%%%%%%%%%%%%%%%%%%

\clearpage
%%%%%%%%%%%%%%%%%%%%%%%%%%%%%%%%%%%%%%%%%%%%%%%%%%%%%%%%%%%%%%%%%%%%%%%%%%%%%%%%

\end{document}